\documentclass[a4paper]{IEEEtran}
\usepackage{cite}
\ifCLASSINFOpdf
  \usepackage[pdftex]{graphicx}
\else
  \usepackage[dvips]{graphicx}
\fi
\usepackage[cmex10]{amsmath}
\usepackage{array}
\usepackage{theorem}
\usepackage[caption=false,font=footnotesize]{subfig}
\usepackage{fixltx2e}
\usepackage{url}
\usepackage{acronym}

{\theorembodyfont{\rmfamily}
   }
{\theorembodyfont{\rmfamily}
   }
{\theorembodyfont{\rmfamily}
   }
{\theorembodyfont{\rmfamily}
   }

{\theorembodyfont{\rmfamily}
   }
{\theorembodyfont{\rmfamily}
   \newtheorem{Exa}{{\textbf Example}}[section]}

\acrodef{LDPC}{low-density parity-check}
\acrodef{MDPC}{moderate-density parity-check}
\acrodef{QC}{quasi-cyclic}
\acrodef{QC-LDPC}{quasi-cyclic low-density parity-check}
\acrodef{QC-MDPC}{quasi-cyclic moderate-density parity-check}
\acrodef{RSA}{Rivest, Shamir, Adleman}
\acrodef{BF}{bit flipping}
\acrodef{SPA}{sum product algorithm}
\acrodef{RDF}{random difference families}
\acrodef{ISDA}{information set decoding attacks}
\acrodef{DCA}{dual code attacks}
\acrodef{WF}{work factor}
\acrodef{BER}{bit error rate}
\acrodef{FER}{frame error rate}
\acrodef{CER}{codeword error rate}
\acrodef{BLER}{block error rate}
\acrodef{PLS}{Physical layer security}
\acrodef{SNR}{signal-to-noise ratio}
\acrodefplural{SNR}[SNRs]{signal-to-noise ratios}
\acrodef{eBCH}{extended Bose-Chaudhuri-Hocquenghem}
\acrodef{CCs}{convolutional codes}
\acrodef{CC}{convolutional code}
\acrodef{UB}{union bound}
\acrodef{TUB}{truncated union bound}
\acrodef{QSFC}{quasi-static fading channel}
\acrodefplural{QSFC}[QSFCs]{quasi-static fading channels}
\acrodef{FFC}{fast fading channel}
\acrodef{CSI}{channel state information}
\acrodef{AWGN}{additive white Gaussian noise}
\acrodef{NMS}{normalized min-sum}
\acrodef{LLR}{log likelihood ratio}
\acrodef{LLR-SPA}{log-likelihood ratio sum product algorithm}
\acrodef{BCC}{broadcast channel with confidential messages}
\acrodef{UEP}{unequal error protection}
\acrodef{BPSK}{binary phase shift keying}
\acrodef{ML}{maximum likelihood}
\acrodef{AONT}{all-or-nothing transform}
\acrodefplural{AONT}[AONTs]{all-or-nothing transforms}
\acrodef{PC}{protection class}
\acrodefplural{PC}[PCs]{protection classes}
\acrodef{MIMO}{multiple-input multiple-output}

\begin{document}

\title{LDPC coded transmissions over the Gaussian broadcast channel with confidential messages
\thanks{This work was supported in part by the MIUR project ``ESCAPADE''
(Grant RBFR105NLC) under the ``FIRB -- Futuro in Ricerca 2010'' funding program.}
}

\author{\IEEEauthorblockN{Marco Baldi, Nicola Maturo, Giacomo Ricciutelli, Franco Chiaraluce,\\}
\IEEEauthorblockA{DII, Universit\`a Politecnica delle Marche,\\
Ancona, Italy\\
Email: \{m.baldi, n.maturo, f.chiaraluce\}@univpm.it}, g.ricciutelli@gmail.com}

\maketitle

\pagestyle{empty}
\thispagestyle{empty}

\begin{abstract}
We design and assess some practical \ac{LDPC} coded transmission schemes for the Gaussian \ac{BCC}.
This channel model is different from the classical wiretap channel model as the unauthorized receiver (Eve) must be able to decode
some part of the information. Hence, the reliability and security targets are different from those of the wiretap channel.
In order to design and assess practical coding schemes, we use the error rate as a metric of the performance achieved
by the authorized receiver (Bob) and the unauthorized receiver (Eve).
We study the system feasibility, and show that two different levels of protection against noise are required on
the public and the secret messages.
This can be achieved in two ways: i) by using \ac{LDPC} codes with \ac{UEP} of the transmitted information bits
or ii) by using two classical non-\ac{UEP} \ac{LDPC} codes with different rates.
We compare these two approaches and show that, for the considered examples, the solution exploiting \ac{UEP} \ac{LDPC} codes 
is more efficient than that using non-\ac{UEP} \ac{LDPC} codes.
\end{abstract}


\section{Introduction}
\label{sec:Intro}

The \ac{BCC} \cite{Csiszar1978} is a well-known transmission model for communications achieving security at the physical layer,
which generalizes Wyner's wiretap channel model \cite{Wyner1975}.
Since its introduction, a lot of work has been done to study the \ac{BCC} from the information theory standpoint, mostly aimed 
at computing the secrecy capacity regions for this channel and its several variants (see \cite{Dijk1997, Liu2009, Chia2012} 
and the references therein).
More recently, the secrecy capacity regions have been studied also for the \ac{BCC} with \ac{MIMO} \cite{Ekrem2012, Liu2013, Bagherikaram2013}
and cooperative communications \cite{Wyrembelski2012}.

For the classical wiretap channel, the use of several practical families of codes has already been investigated:
this is the case of lattice codes \cite{Belfiore2010}, polar codes \cite{Mahdavifar2010} and \ac{LDPC} codes \cite{Klinc2011}.
Instead, for the \ac{BCC}, despite the large amount of theoretical work, there is still a lack of practical systems 
able to achieve some specific security and reliability targets.
The use of coding is recognized as an important tool also in such a context, but most studies consider the abstraction of 
random coding \cite{Watanabe2012}, which indeed is difficult to translate into a practical coding scheme.
At the authors' best knowledge, the only proposal of using a family of practical codes over this special channel 
appeared very recently in \cite{Andersson2013}, and exploits polar codes.
Other, and even more widespread families of codes, like \ac{LDPC} codes, have never been considered in such a context.

In this paper, we focus on the Gaussian \ac{BCC} and study some practical \ac{LDPC} coded transmission schemes
for achieving reliability and security over this channel.
For this purpose, we follow some recent literature and use the error rate as a metric \cite{Klinc2011, WongWong2011, Baldi2010, Baldi2011, Baldi2012}.
We define suitable reliability and security targets for the Gaussian \ac{BCC} in terms of the error rate, 
and redefine the concept of \textit{security gap}, defined for the Gaussian wiretap channel as the quality ratio between 
Bob's and Eve's channels needed to achieve the reliability and security targets.

We consider \ac{LDPC} codes, since they are state-of-the-art codes able to approach the channel capacity under iterative decoding.
We show that, in order to achieve transmission reliability and security over the \ac{BCC}, a coding scheme with two
different levels of protection against noise is needed.
For this reason, we consider an \ac{LDPC} code with \ac{UEP} capability, and compare its performance 
with that achievable by using two different non-\ac{UEP} \ac{LDPC} codes.

The organization of the paper is as follows:
in Section \ref{sec:Model} we define the system model and the metrics adopted.
In Section \ref{sec:SingleCodes} we study the use of single codes with different rates.
In Section \ref{sec:UEPcodes} we introduce \ac{UEP} \ac{LDPC} codes into the system.
In Section \ref{sec:Comparison} we assess the performance achievable through the considered codes and
Section \ref{sec:Conclusion} concludes the paper.

\section{System model}
\label{sec:Model}

In the Gaussian \ac{BCC}, we have one transmitter (Alice) sending broadcast and confidential information over
the channel.
Bob is able to decode the whole information, while Eve is able to get only the public message, ideally without gathering
any useful information on the secret message.
Both the Alice-Bob and the Alice-Eve channels are supposed to be Gaussian. 

We assume that each transmitted message is formed by $n$ bits and includes a public and a confidential part.
We also suppose to use coding, and that each transmitted message contains $k$ information bits and $r=n-k$ redundancy bits.
It follows that the overall code rate is $R = \frac{k}{n}$, and $R$ also coincides with the overall information rate,
expressed in bits per channel use, under the hypothesis of \ac{BPSK} modulation.
In our model, each transmitted message contains a block of $k_s \le k$ information bits which are secret, 
while the remaining $k_p = k - k_s$ information bits form a block of public information.
It follows that the secret and public information rates are $R_s = \frac{k_s}{n}$ and $R_p = \frac{k_p}{n}$, respectively,
and $R = R_s + R_p$.

Concerning the redundancy part, we can suppose that it can be split into two groups: $r_s \le r$ 
redundancy bits are used to check the $k_s$ secret information bits, while the remaining $r_p = r - r_s$ bits check 
the public information bits.
This hypothesis will be removed when we will consider codes with \ac{UEP}, in which some protection
classes are defined without splitting the redundancy among them.
If we assume to use two different channel codes for the secret and the public parts, their code rates are
$R_c^{(s)} = \frac{k_s}{k_s + r_s}$ and $R_c^{(p)} = \frac{k_p}{k_p + r_p}$, respectively.
If we define $\rho = \frac{k_s + r_s}{n}$, we have $R_s = R_c^{(s)} \rho$, $R_p = R_c^{(p)} (1-\rho)$ and
$R = R_c^{(s)} \rho + R_c^{(p)} (1-\rho)$.

\subsection{Reliability and security metrics}
\label{subsec:Metrics}

We consider that both Bob's and Eve's channels are \ac{AWGN} channels or, equivalently, \acp{QSFC}
with channel gains $\gamma^{(B)}$ and $\gamma^{(E)}$, respectively, expressed in \ac{SNR} per bit.
Other channel models, like the fast fading channel, are outside the scope of this paper, and will be studied
in future works.
$P(\gamma)$ denotes the overall \ac{FER} as a function of the \ac{SNR} $\gamma$,
that is, the probability that, within a received frame of $n$ bits, one or more of the 
$k$ information bits are in error after decoding.
Similarly, $P_s(\gamma)$ ($P_p(\gamma)$) denotes the \ac{BLER} for
the secret (public) information block, i.e., the probability that, within a received 
frame of $n$ bits, one or more of the $k_s$ ($k_p$) secret (public) information bits 
are in error after decoding.
Let us fix two small threshold values, $\delta$ and $\eta$, and define the security and reliability
targets in terms of the decoding error probability as follows:
\begin{subequations}
\begin{align}
P_p(\gamma^{(B)}) & \le \delta, \label{eq:Pconda} \\
P_p(\gamma^{(E)}) & \le \delta, \label{eq:Pcondb} \\
P_s(\gamma^{(B)}) & \le \delta, \label{eq:Pcondc} \\
P_s(\gamma^{(E)}) & \ge 1 - \eta. \label{eq:Pcondd} 
\end{align}
\label{eq:Pconditions}
\end{subequations}

\begin{figure}[!t]
\begin{centering}
\includegraphics[width=60mm,keepaspectratio]{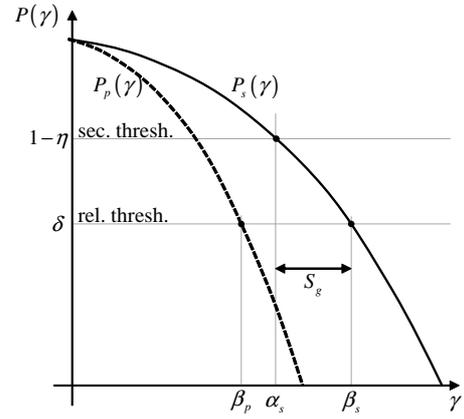}
\caption{Expected block error rate curves for the public and secret messages as functions of the \ac{SNR}.}
\label{fig:ErrorRatePlot}
\par\end{centering}
\end{figure}

Let us suppose that the public information blocks are more protected against noise than 
the secret information blocks.
This scenario is exemplified in Fig. \ref{fig:ErrorRatePlot}, where we suppose that the public 
information blocks experience a lower \ac{BLER} than the secret information blocks.
Conditions \eqref{eq:Pconditions} can then be translated in terms of Bob's and Eve's \acp{SNR}, 
i.e., $\gamma^{(B)}$ and $\gamma^{(E)}$, respectively.
More precisely, by looking at the figure, we have that conditions \eqref{eq:Pconda} and \eqref{eq:Pcondc}
become
\begin{equation}
\gamma^{(B)} \ge \max\left\{\beta_p, \beta_s\right\} = \beta_s,
\label{eq:Bobcond}
\end{equation}
whereas conditions \eqref{eq:Pcondb} and \eqref{eq:Pcondd} become
\begin{equation}
\beta_p \le \gamma^{(E)} \le \alpha_s.
\label{eq:Evecond}
\end{equation}

It follows from \eqref{eq:Evecond} that, for the system to be feasible, we must actually ensure that the
public message is more protected against noise than the secret one (this typically implies $R_c^{(p)} < R_c^{(s)}$).
In fact, if the opposite occurs, since $1-\eta > \delta$, we have $\alpha_s < \beta_p$, and condition \eqref{eq:Evecond} cannot be met.
From the theoretical standpoint, the system is feasible even when $\alpha_s = \beta_p$.
This obviously is a limit condition, while from the practical standpoint it is useful that $\alpha_s > \beta_p$, and that the ratio 
$\frac{\alpha_s}{\beta_p}$ is quite greater than one, such that the system remains feasible even when $\gamma^{(E)}$ has some fluctuations.
In this work we neglect this fact, since we only consider static (or quasi-static) channels, and the only constraint we impose is $\alpha_s \ge \beta_p$.
The ratio $\frac{\alpha_s}{\beta_p}$ will be studied in future works, where non-static channels will be considered as well.

When the system is feasible, i.e., the public message is more protected against noise than the secret one,
and $\alpha_s \ge \beta_p$, we can compare different coding techniques by using 
the security gap $S_g$, defined as the ratio between Bob's minimum \ac{SNR}
and Eve's maximum \ac{SNR}:
\begin{equation}
S_g = \frac{\beta_s}{\alpha_s}.
\label{eq:Sg}
\end{equation}
Obviously, the smaller the security gap, the better the system performance, since security can be achieved
even with a small degradation of Eve's channel with respect to Bob's channel.

Based on the above considerations, the design target is to find codes which make the system feasible.
In fact, differently from the wiretap channel model, in this case there is no guarantee that the system
is feasible even when Eve has a degraded channel with respect to Bob.
Then, a meaningful objective is to find codes able to achieve small security gaps.
We will face these problems in the next sections.

\subsection{Message concatenation and all-or-nothing transforms}
\label{subsec:ConcatAONT}

In order to increase the difference between the two levels of protection against noise for the public and secret
messages, we can resort to message concatenation \cite{Baldi2012} and \acp{AONT} \cite{Boyko1999}.
Let us suppose that $L$ secret messages, each with length $k_s$, are concatenated and
then transformed through an \ac{AONT}. The transformed string is then transmitted in $L$ fragments,
which replace the original messages.
Only if all of them are correctly received, the \ac{AONT} can be inverted and the $L$ secret messages 
successfully obtained; otherwise, none of them can be even partially recovered.
Through concatenation, the error probability on each secret message becomes
\begin{equation}
P_s^{(L)}(\gamma) = 1 - \left[ 1 - P_s(\gamma) \right]^L \ge P_s(\gamma).
\end{equation}

Hence, for a given $\gamma^{(E)} = \bar{\gamma}^{(E)}$, if $P_s(\bar{\gamma}^{(E)})$ does not meet the
security condition, we can resort to message concatenation and \acp{AONT}, and find a
suitable value of $L$ such that $P_s^{(L)}(\bar{\gamma}^{(E)})$ overcomes the security threshold.

Obviously, when we introduce message concatenation and \acp{AONT}, we must replace $P_s(\gamma)$
with $P_s^{(L)}(\gamma)$ also for Bob. Hence, the use of these tools is paid in terms of the \ac{SNR}
working point for Bob, which increases with respect to the case without concatenation.
In addition, increasing $L$ increases the latency for receiving the secret message.
Concerning the implementation of an \ac{AONT}, several examples can be found in the literature.
For the purposes of this study, we observe that scrambling the information bits through a linear (and dense) 
map can achieve features similar to those of an \ac{AONT}, thanks to the randomness of the errors
induced by the channel \cite{Baldi2012}.

We note that \acp{AONT} can also be used, at higher layers, to achieve some desired level
of computational security.
In fact, the condition \eqref{eq:Pcondd} only guarantees that Eve's decoder has a high error probability
on the secret blocks. However, this does not exclude that some secret blocks may be correctly decoded by Eve.
Furthermore, even when Eve's decoder is in error, some bits within the block may be correct.
Therefore, as often occurs in physical layer security, this setting represents a substrate which must
be exploited by higher layer protocols to achieve some desired level of computational security.

\section{Using two different \ac{LDPC} codes}
\label{sec:SingleCodes}

Let us suppose to use two different \ac{LDPC} codes to encode the public and the secret information blocks.
For the sake of simplicity, our choice is to split the transmitted frame into two codewords of length $n/2$.
One of these two codewords is obtained from an \ac{LDPC} code $C_p$, having rate $R_c^{(p)}$, 
and carries the $k_p$ public information bits. 
The other codeword belongs to an \ac{LDPC} code $C_s$, with rate $R_c^{(s)}$ and corresponds
to the $k_s$ secret information bits.
Since the two codes have the same length, provided that they are well designed, it must be $R_c^{(p)} < R_c^{(s)}$ to achieve a higher level of protection
against noise for the public information block.

\begin{Exa}
\label{exa:SingleCodes}
Let us consider $n=2048$ and two \ac{LDPC} codes with the following parameters:
\begin{itemize}
\item $C_p$: length $1024$, rate $R_c^{(p)}=0.2$.
\item $C_s$: length $1024$, rate $R_c^{(s)}=0.8$.
\end{itemize}
Their variable and check node degree distributions have been optimized through the tools available in \cite{LODE2013}.
Concerning the choice of the node degrees, for the variable nodes we have used the same degrees we will consider
in Example \ref{exa:UEPCodes}, while for the check nodes we have considered a concentrated distribution 
(i.e., with only two degrees, concentrated around the mean).
The resulting variable and check node degree distributions are, respectively,
\begin{align}
\lambda(x) & = 0.1765 x^{19} +	0.2392 x^{18} + 0.0638 x^{17} + 0.0988 x^{16} \nonumber \\
							 & + 0.0117 x^{15} + 0.1976 x^2 + 0.2124 x, \nonumber \\
\rho(x) & = 0.1607 x^6 + 0.8393 x^5,
\end{align}
for the first code, and
\begin{align}
\lambda(x) & = 0.8815 x^2 +	0.1185 x, \nonumber \\
\rho(x) & = 0.1708 x^{14} + 0.8292 x^{13},
\end{align}
for the second code.
These degree distributions have been used to design the parity-check matrices of the two codes
$C_p$ and $C_s$ through the \textit{zigzag-random} construction \cite{Hu2001PEG, Deetzen2010}.
The performance of these codes, assessed through numerical simulations, and using the \ac{LLR-SPA}
with $100$ maximum iterations for decoding, is reported in Fig. \ref{fig:SingleCodes}, also considering some examples of concatenation
of the secret message ($L=100,1250,10000$).
\begin{figure}[!t]
\begin{centering}
\includegraphics[width=70mm,keepaspectratio]{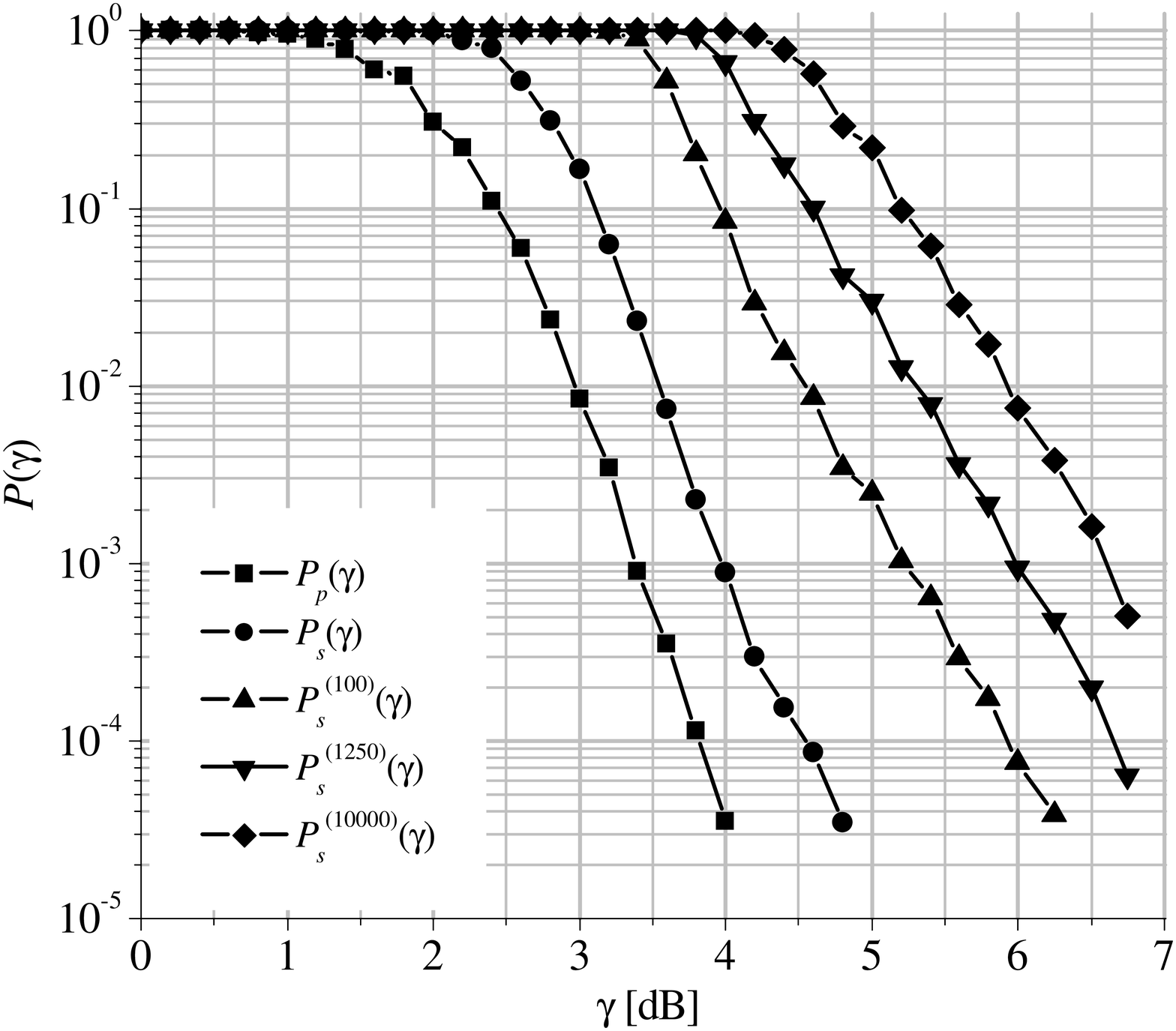}
\caption{Error rate curves for two different \ac{LDPC} codes with length $n = 1024$ and rates $R_c^{(p)}=0.2, R_c^{(s)}=0.8$, with and without concatenation of the secret messages
(indicated in the superscript of $P_s(\gamma)$).}
\label{fig:SingleCodes}
\par\end{centering}
\end{figure}
\end{Exa}

\section{Using \ac{UEP} \ac{LDPC} codes}
\label{sec:UEPcodes}

Let us suppose to use a single \ac{UEP} \ac{LDPC} code with length $n$.
Most of the existing works on \ac{UEP} \ac{LDPC} codes aim at designing codes with
three \acp{PC}:
\begin{itemize}
\item \ac{PC}1 contains $k_1 < k$ information bits which are those most protected against noise.
\item \ac{PC}2 contains $k_2 = k - k_1$ information bits which are less protected against noise than those in \ac{PC}1.
\item \ac{PC}3 contains the whole redundancy part ($r = n-k$ bits).
\end{itemize}
Codes of this kind are suitable for the considered scenario.
In fact, given an \ac{UEP} \ac{LDPC} code with the three \acp{PC} outlined above, we
can map the public message bits into \ac{PC}1 (i.e., $k_p = k_1$) and the secret message
bits into \ac{PC}2 (i.e., $k_s = k_2$).

To design \ac{LDPC} codes with good \ac{UEP} properties, several approaches have been
proposed in the literature \cite{Deetzen2010, Neto2011, Poulliat2007}.
All these methods aim at optimizing the node degree distributions in such a way that
the variable node degrees are spanned in a wide range, and good convergence thresholds
are achieved under iterative decoding.
Then the variable nodes with the highest degrees are mapped into the bits of \ac{PC}1,
whereas the others form \ac{PC}2 and \ac{PC}3 (depending on their association with information
or redundancy bits).

Once the variable node degree distribution $\lambda(x)$ has been designed, the number of bits 
in \ac{PC}1 can be easily computed by converting $\lambda(x)$ from the edge perspective to the node perspective,
the latter being expressed through another polynomial $\nu(x) = \sum_i \nu_i x^i$, and then computing the fraction of variable nodes with the highest degrees,
that are those in \ac{PC}1.
We have
\begin{equation}
\nu_i  = \frac{\lambda_i/ i}{\sum_{j=1}^{\overline{d_v}} \lambda_j/j}, \, \, \, \lambda_i  = \frac{\nu_i \cdot i}{\sum_{j=1}^{\overline{d_v}} \nu_j \cdot j},
\end{equation}
where $\overline{d_v}$ denotes the maximum variable node degree.
The same formulas can also be used for the check node degree distributions, by putting $\rho$ in place of $\lambda$, $c$ in place of $\nu$ and $\overline{d_c}$ in place of $\overline{d_v}$, where $\overline{d_c}$ is the maximum check node degree. Hence, $\rho(x)$ and $c(x)$ are the check node degree distributions from the edge and the node perspectives, respectively.

For the sake of simplicity, for the check node degrees we adopt a concentrated distribution,
as already done in Section \ref{sec:SingleCodes} for the case of different LDPC codes.
Hence, we have 
\begin{equation}
c(x) = a x^{\left\lfloor c_m \right\rfloor} + b x^{\left\lceil c_m \right\rceil},
\end{equation}
where $c_m = \frac{E}{r} = \frac{\sum_{j} v_j \cdot j}{(1-R)}$ and $E$ is the total number 
of edges in the Tanner graph.
The coefficients $a$ and $b$ are obtained as:
\begin{equation}
a = \lceil c_m \rceil - c_m, \ \ \ \ b = c_m - \lfloor c_m \rfloor.
\end{equation}

\begin{Exa}
\label{exa:UEPCodes}
Let us consider the following \ac{UEP} \ac{LDPC} variable node degree distribution taken
from \cite[Table 3]{Poulliat2007}, with some minor modifications to adapt the proportion
between \ac{PC}1 and \ac{PC}2 in such a way that it coincides with the one used in Example \ref{exa:SingleCodes}:
\begin{align}
\lambda (x) & = 0.0025 x^{19} + 0.0009 x^{18} + 0.0031 x^{17} + 0.0630 x^{16} \nonumber \\
                & + 0.3893 x^{15} + 0.2985 x^{2} + 0.2427 x.
\end{align}
The corresponding node perspective distribution is
\begin{align}
\nu (x) & = 0.0005 x^{20} + 0.0002 x^{19} + 0.0007 x^{18} + 0.0151 x^{17} \nonumber \\
          & + 0.0835 x^{16} + 0.4054 x^{3} + 0.4946 x^{2}.
\end{align}
\begin{figure}[!t]
\begin{centering}
\includegraphics[width=70mm,keepaspectratio]{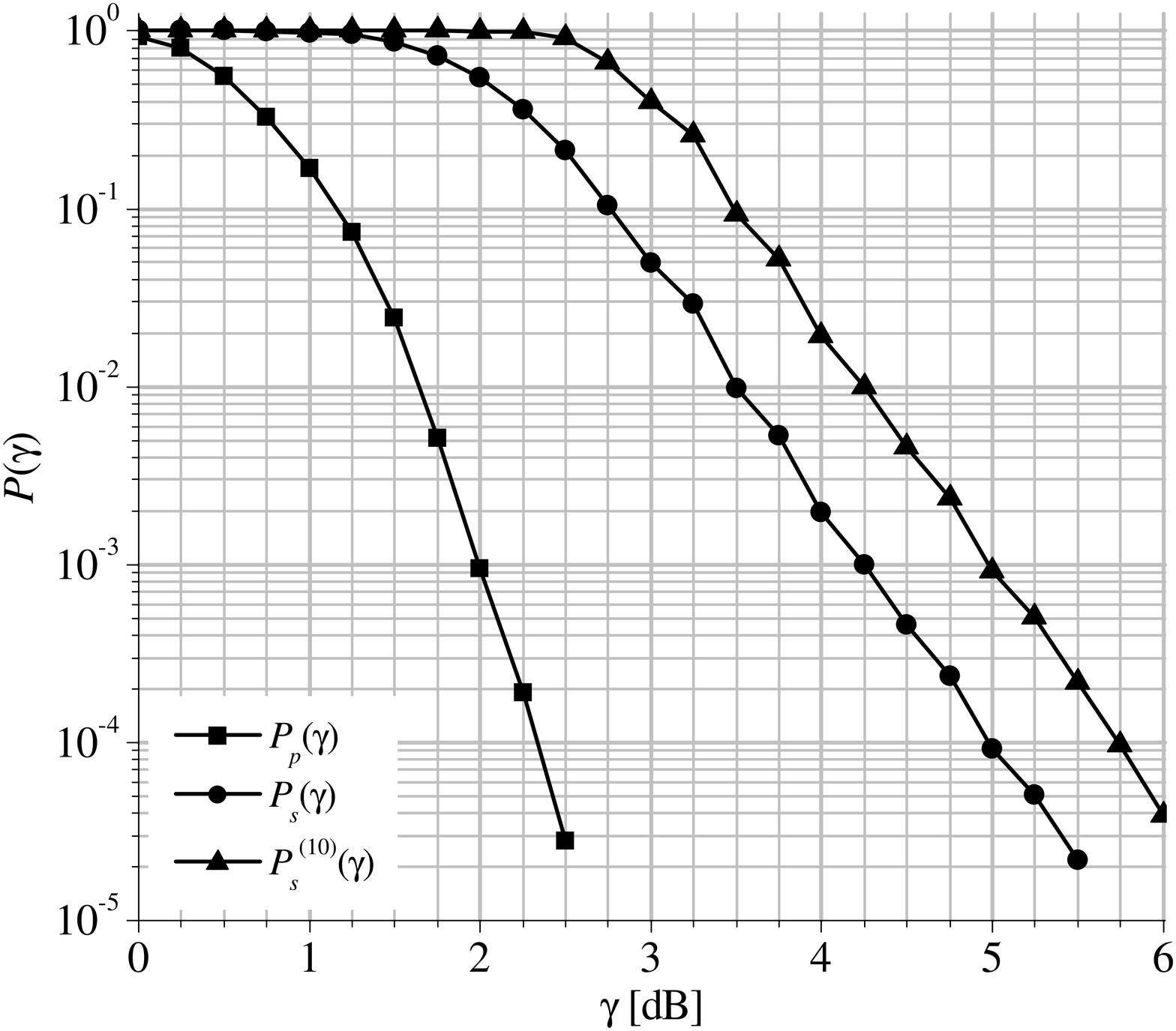}
\caption{Error rate curves for an \ac{UEP} \ac{LDPC} code with length $n = 1024$ and \ac{PC}1 and \ac{PC}2 with proportions $20\%-80\%$, with and without concatenation of secret messages
(indicated in the superscript of $P_s(\gamma)$).}
\label{fig:UEP1024}
\par\end{centering}
\end{figure}
\begin{figure}[!t]
\begin{centering}
\includegraphics[width=70mm,keepaspectratio]{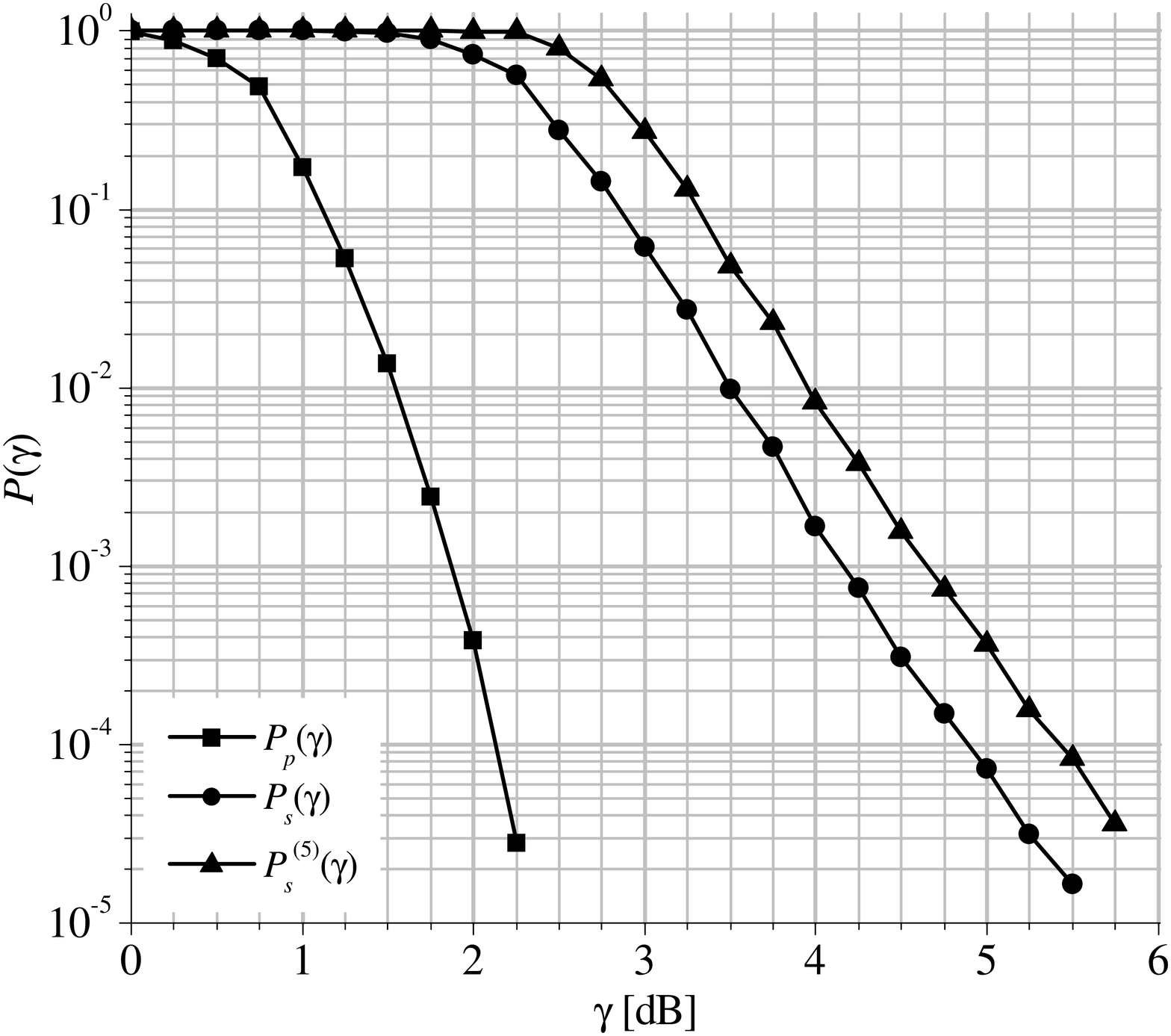}
\caption{Error rate curves for an \ac{UEP} \ac{LDPC} code with length $n = 2048$ and \ac{PC}1 and \ac{PC}2 with proportions $20\%-80\%$, with and without concatenation of secret messages
(indicated in the superscript of $P_s(\gamma)$).}
\label{fig:UEP2048}
\par\end{centering}
\end{figure}
\begin{figure}[!t]
\begin{centering}
\includegraphics[width=70mm,keepaspectratio]{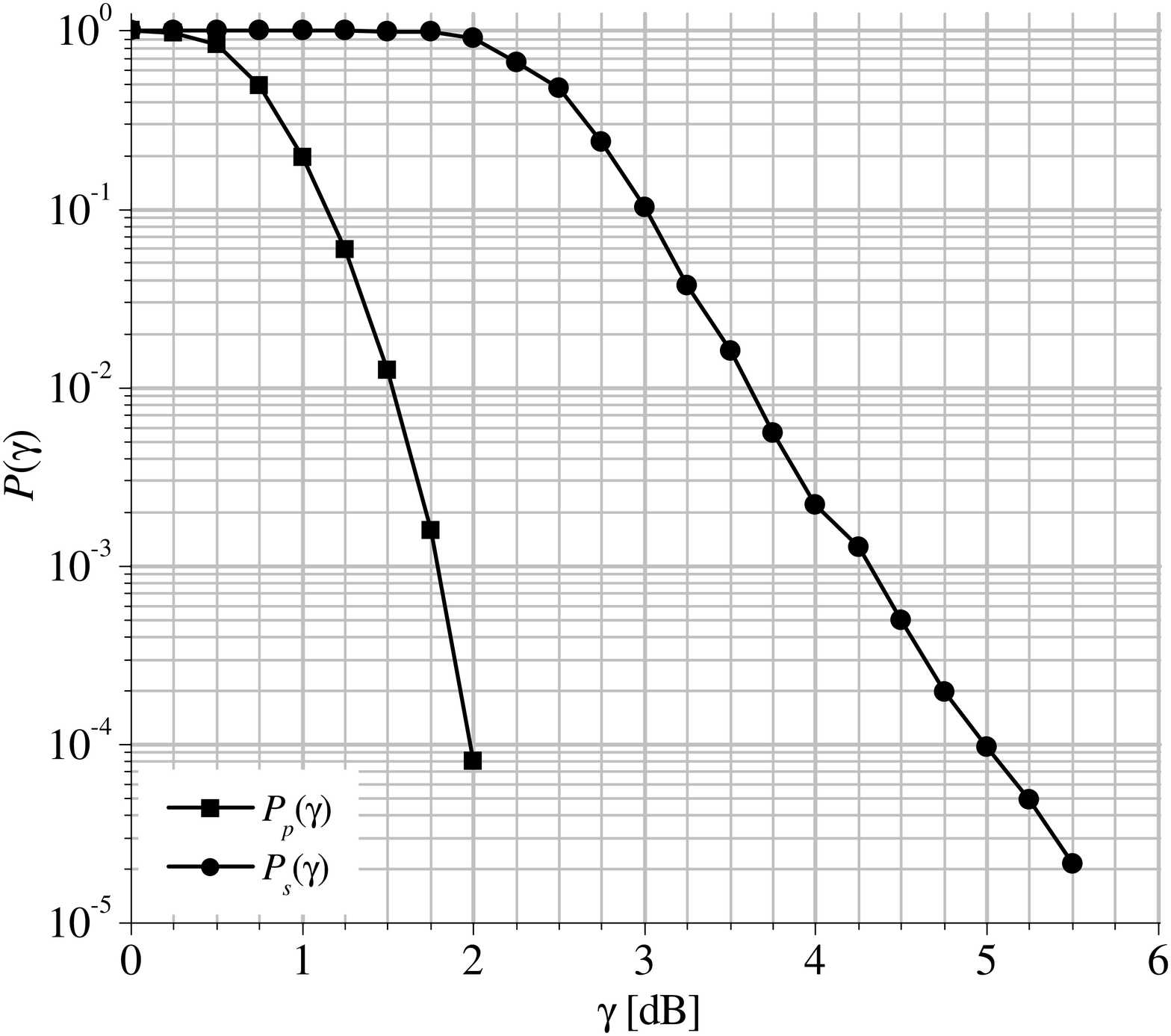}
\caption{Error rate curves for an \ac{UEP} \ac{LDPC} code with length $n = 4096$ and \ac{PC}1 and \ac{PC}2 with proportions $20\%-80\%$.}
\label{fig:UEP4096}
\par\end{centering}
\end{figure}

The nodes in \ac{PC}1 are those with degree $\ge 16$, while those with degree $\le 3$ are in \ac{PC}2 or \ac{PC}3 depending on their association to information bits or redundancy bits. This way, we find that \ac{PC}1 and \ac{PC}2 contain, respectively, $20\%$ and $80\%$ of the information bits.
By using this distribution for the variable nodes and a concentrated degree distribution for the check nodes, we have
designed three \ac{UEP} \ac{LDPC} codes with $n=1024, 2048$ and $4096$.
Their parity-check matrices have been obtained through the same zigzag random procedure used
in Section \ref{sec:SingleCodes}.
The performance obtained by these codes under \ac{LLR-SPA} decoding with $100$ maximum iterations is reported in Figs. \ref{fig:UEP1024}-\ref{fig:UEP4096}.
Some examples of the use of concatenation of secret messages are also shown in Figs. \ref{fig:UEP1024} and \ref{fig:UEP2048}.

\end{Exa}

\section{Performance assessment}
\label{sec:Comparison}

We fix two values for the reliability and security thresholds, namely, $\delta = 10^{-4}$ and $\eta = 0.1$.
Actually, one could think that a decoding error probability equal to $0.9$ for Eve does not
represent a condition of sufficient security.
However, we remind that this setting only provides a substrate over which any desired level of computational security
can be achieved through higher layer techniques, as described in Section \ref{subsec:ConcatAONT}.
Furthermore, our purpose is just to compare the considered coding schemes, not to define any absolute security level.
For each coding scheme, we choose the smallest value of $L$ such that the system is feasible, \textit{i.e.}, $\alpha_s \ge \beta_p$.
Finally, we compute the values of $\beta_s$ and the security gap $S_g$, according to \eqref{eq:Sg}.

The results obtained by considering the coding schemes in Examples \ref{exa:SingleCodes} and \ref{exa:UEPCodes} are reported in Table \ref{tab:Comparison}.
From these examples, we observe that using \ac{UEP} \ac{LDPC} codes is actually effective for implementing practical transmission schemes
over the \ac{BCC}, since the system feasibility is achieved even with a small number of concatenated messages,
and the security gap values are in the order of $3 - 3.3$ dB.
Increasing the block length improves performance:
apart from a small reduction in the security gap, longer codes require a smaller \ac{SNR} for Bob and less concatenation.
In fact, while an \ac{UEP} \ac{LDPC} code with $n=1024$ requires $L=10$
and $\beta_s = 5.74$ dB, by increasing $n$ to $4096$ we reduce $\beta_s$ to less than $5$ dB (thus reducing Bob's \ac{SNR}),
and we no longer need the concatenation of secret messages for the system to be feasible.
Instead, using two different codes is not a good choice, as we observe by comparing the second and the third rows
of Table \ref{tab:Comparison}.
In fact, for $n=2048$, the two non-\ac{UEP} \ac{LDPC} codes considered in Example \ref{exa:SingleCodes}
achieve some small reduction in the security gap, but they require
a very high level of concatenation ($L=1250$) for the system to be feasible.
This increases the minimum \ac{SNR} for Bob by more than $1$ dB, and also has detrimental effects on the system latency.
\begin{table}[!t]
\renewcommand{\arraystretch}{1.1}
\caption{Performance assessment of the coding schemes in Examples \ref{exa:SingleCodes} and \ref{exa:UEPCodes}  ($\beta_p, \alpha_s, \beta_s$ and $S_g$ are in $\mathrm{dB}$) .}
\label{tab:Comparison}
\centering
\begin{tabular}{|c|c|c|c|c|c|c|}
\hline
Scheme & $n$ & $L$ & $\beta_p$ & $\alpha_s$ & $\beta_s$ & $S_g$ \\
\hline
\hline
\ac{UEP} & $1024$ & $10$ & $2.34$ & $2.46$ & $5.74$ & $3.28$ \\
\hline
non-\ac{UEP} & $2048$ & $1250$ & $3.81$ & $3.83$ & $6.65$ & $2.82$ \\
\hline
\ac{UEP} & $2048$ & $5$ & $2.13$ & $2.37$ & $5.43$ & $3.06$ \\
\hline
\ac{UEP} & $4096$ & $1$ & $1.99$ & $2.01$ & $4.98$ & $2.97$ \\
\hline
\end{tabular}
\end{table}

\section{Conclusion}
\label{sec:Conclusion}

We have studied the performance of some practical \ac{LDPC} coded transmission schemes for the \ac{BCC}.
We have used the error rate as a metric, and proved that two different levels of protection against noise
are needed for the public and the secret messages.

For this purpose, we have considered both \ac{UEP} \ac{LDPC} codes and classical non-\ac{UEP} \ac{LDPC} codes.
We have considered some specific sets of parameters to provide some practical examples.
For the considered cases, our results show that rather small security gaps can be achieved, and that using long \ac{UEP} \ac{LDPC} codes is advantageous,
since it allows to avoid the use of message concatenation, thus reducing the required \ac{SNR} and the transmission latency.


\newcommand{\BIBdecl}{\setlength{\itemsep}{0.005\baselineskip}}
\bibliographystyle{IEEEtran}
\bibliography{Archive}

\end{document}